\documentclass[useAMS,usenatbib,usegraphicx]{mn2e}
\usepackage{amsmath,amssymb}

\title[A wide-field survey of globular clusters - II]{A {\it
Subaru/Suprime-Cam} wide-field survey of globular cluster populations
around M87 - II: Colour and spatial distribution}

\author[N. Tamura et al.]{Naoyuki
Tamura$^{1}$\thanks{E-mail:naoyuki@naoj.org}\thanks{Current address:
Subaru Telescope, National Astronomical Observatory of Japan,
650 North A'ohoku Place, Hilo, HI 96720, USA}, Ray
M. Sharples$^{1}$, Nobuo Arimoto$^{2}$, Masato Onodera$^{2, 3}$,
\newauthor
Kouji Ohta$^{4}$, \& Yoshihiko Yamada$^{2}$ \\
$^{1}$Department of Physics, University of Durham, South Road, Durham,
DH1 3LE, United Kingdom \\
$^{2}$National Astronomical Observatory of Japan, Mitaka, Tokyo
181-8588, Japan \\
$^{3}$Department of Astronomy, School of Science, University of Tokyo,
Tokyo 113-0033, Japan \\
$^{4}$Department of Astronomy, Faculty of Science, Kyoto University,
Kyoto 606-8502, Japan
}
 
\begin{document}

\date{}
  
\pagerange{\pageref{firstpage}--\pageref{lastpage}} \pubyear{2006}
  
\maketitle

\label{firstpage}
  
\begin{abstract}

We have performed a wide-field imaging survey of the globular cluster
(GC) populations around M87 with Suprime-Cam on the 8.2m Subaru
Telescope. A $2^{\circ} \times 0_{\cdot}^{\circ}5$ field extending from
the centre of M87 out to a radius of $\sim$ 0.5 Mpc was observed through
the $BVI$ filters.
By investigating the GC colour distribution as a function of distance
from M87 and NGC 4552, another luminous Virgo elliptical in our survey
field, it is found that clear bimodality ($(V-I)_{\rm peak} \sim 1.0$
and 1.2) is seen only in the inner regions ($\lesssim 10$ kpc) of the
host galaxies and that it becomes weaker with radius due to the
decreasing contribution of the red GC ($V-I > 1.1$) subpopulation.
It is also found (both around M87 and NGC 4552) that while the spatial
distribution of the red GCs is as centrally concentrated as the host
galaxy halo light distribution, the distribution of the blue GCs ($V-I
\leq 1.1$) tends to be more extended. However, the blue GC distribution
around M87 is not as extended as the Virgo cluster mass density
profile. Based on these facts, we argue that the contribution of GCs
associated with the Virgo cluster (e.g., intergalactic GCs) is not
significant around M87 and is not the primary reason for the high $S_N$
value of M87. Instead, most of the blue GCs around luminous ellipticals,
as well as the red GCs, are presumed to be associated with the host
galaxy.

We model the radial profile of GC surface densities out to $\sim$ 0.5
Mpc from M87 by a superposition of the GC populations associated with
M87 and with NGC 4552. It is found that there are some regions where the
GC surface densities are larger
than expected from this model, suggesting the existence of an additional
intergalactic GC (i-GC) population independent of the luminous
ellipticals.
By comparing the estimated i-GC surface density with the luminosity
density of the intracluster stellar population inferred from the
intergalactic planetary nebulae in the Virgo cluster, we obtain a crude
estimate of i-GC specific frequency $S_N = 2.9^{+4.2}_{-1.5}$. If this
$S_N$ value represents the stellar population tidally stripped by a
massive central galaxy from other (less luminous) galaxies, the
contribution of tidally captured GCs in the GC population of M87 would
need to be low to be consistent with the high $S_N$ value of M87.

\end{abstract}


\section{INTRODUCTION}\label{intro}

Globular cluster (GC) populations are one of the key probes of the
archeology of their host galaxy
because GCs contain stars with a single age and metallicity and are
therefore in principle simpler to interpret than the integrated light of
field stars. One of the most exciting recent developments in the study
of extragalactic GCs is the discovery that many luminous elliptical
galaxies have bimodal or multimodal colour distributions of GCs (e.g.,
Gebhardt \& Kissler-Patig 1999; Larsen et al. 2001; Brodie et al.
2005). Since this is unlikely to be produced in a model of elliptical
galaxy formation with a simple monolithic collapse and subsequent star
formation, more
elaborate models have been proposed. One is the multiphase collapse
scenario (Forbes, Brodie \& Grillmair 1997), where all GCs are presumed
to form coevally with the host galaxy in a massive burst of star
formation within a short time scale at high redshift and are therefore
expected to be uniformly old. Since the starburst is presumed to be
split into several discrete phases, some of the GCs will form in gas
clouds more polluted by earlier star formation, which could result in a
red GC subpopulation. Another proposed model is the merger scenario
(Ashman \& Zepf 1992), where luminous ellipticals are presumed to have
formed via gas-rich galaxy mergers which induce the formation of
additional red GCs, while the blue ones originate in the progenitor
galaxies. In this case, the blue and red GC subpopulations would contain
predominantly old/metal-poor GCs and younger/metal-rich GCs,
respectively. Beasley et al. (2002) found that GC bimodality was a
general characteristic of hierarchical merging models in which the
formation of blue GCs was truncated at high redshift.

However, most of the elliptical galaxies whose GC populations have so
far been intensively studied reside in high density environments. The GC
properties upon which the proposed scenarios are based may therefore not
be intrinsic but may have been substantially modified by environmental
effects. Field stars in the outer parts of galaxies can be stripped off
into intergalactic space due to tidal interactions between galaxies and
the cluster potential, and as a result of high-speed encounters between
galaxies. In fact, observations have now provided direct evidence for
such an intracluster stellar population in the form of planetary nebulae
(PNe) (e.g., Theuns \& Warren 1997; Feldmeier et al. 2004a; Arnaboldi et
al. 2004), red giant branch stars (Durrell et al. 2002), and diffuse
light (e.g., Feldmeier et al. 2002, 2004b). It is believed that GCs may
also be stripped from galaxies by tidal interactions, suggesting that
intergalactic space may harbour a potential reservoir of GCs which needs
to be taken into account to understand the evolution of GC populations.
In an alternative explanation of the bimodality of GCs known as the
accretion scenario (C\^{o}t\'{e}, Marzke, \& West 1998), the blue
(metal-poor) GCs are assumed to have been captured from other
(presumably less luminous) galaxies through tidal stripping and/or
accretion.
A population of intergalactic GCs (``i-GC'' hereafter) may naturally
explain the fact that GCs tend to be extremely populous around central
cluster galaxies, because they would be most likely to be surrounded by
a significant population of i-GCs (West et al. 1995).
By surveying directly for i-GCs and characterizing this population, it
is possible to test the validity of the accretion scenario and improve
our understanding of the formation of GC populations. Furthermore, the
i-GC specific frequency could be estimated by comparing the luminosity
of the intracluster stellar population with the total number of
i-GCs. Since GC specific frequency is considered to depend on galaxy
morphology, this would give an insight into the parent galaxy type and
luminosity of i-GCs and the physical processes of galaxy interactions.

To date, however, there is no firm evidence for or against the existence
of i-GCs (Bassino et al. 2003; Jord\'{a}n et al. 2003; Mar\'{i}n-Franch
\& Aparicio 2003). In fact, several technical difficulties must be
overcome to find i-GCs. Firstly, the surface number density of i-GCs is
inferred to be highest around the cluster centre because tidally
stripped GCs are expected to be distributed like the mass distribution
of the host {\it galaxy cluster}
(White 1987; Muzzio 1987; West et al. 1995), but a cD galaxy or giant
elliptical is located at the cluster centre in many cases. Since
luminous ellipticals are normally surrounded by thousands of GCs, a
large area needs to be surveyed to see whether there are additional
i-GCs and to discriminate them from GCs associated with the central
galaxy. This search needs to extend over several hundred kilo-parsecs
from the cluster centre, since the GC population in a luminous
elliptical galaxy may extend out to $\sim$ 100 kpc (Rhode \& Zepf 2001;
Dirsch et al. 2003). By surveying a contiguous region around the central
galaxy, one may be able to detect an apparent change in the slope of the
radial profile of GC surface number densities due to the presence of
i-GCs. Secondly, GCs need to be probed down to a very low surface number
density to investigate the spatial structure of their distribution out
to such large radii. It is suggested by recent wide-field studies of GC
populations in luminous ellipticals that the GC surface number density
in the outer halos is $\lesssim$ 1 arcmin$^{-2}$ (Rhode \& Zepf 2001,
2004; Dirsch et al. 2003; Dirsch, Schuberth \& Richtler 2005). If this
is a mixture of GCs associated with the host galaxy and i-GCs, the i-GC
surface number density would be even lower.  In fact, a recent study of
GC populations in the Fornax cluster has suggested an i-GC surface
number density of $\sim 0.1$ arcmin$^{-2}$ (Bassino et al. 2003).
Accordingly, selection of GCs and statistical subtraction of foreground
and background contamination will have to be carefully performed.

In this paper and a companion paper (Tamura et al. 2006; Paper I
hereafter), we report a wide-field imaging survey in the $B$, $V$, and
$I$ bands of the GC populations around M87, the luminous elliptical
galaxy at the centre of the Virgo cluster, conducted with Suprime-Cam on
the Subaru Telescope. The survey area is $\sim 2^{\circ} \times
0_{\cdot}^{\circ}5$ (560 kpc $\times$ 140 kpc) which is the widest
survey yet undertaken of the GC populations around luminous ellipticals
(the survey depth is similar or slightly deeper than the previous
ground-based surveys). We also make use of wide-field imaging data on
control fields taken with Subaru/Suprime-Cam to correct for foreground
and background contamination.
The observations, data reduction and data analyses including the
selection of GC candidates, incompleteness corrections, and subtraction
of foreground and background contamination in the GC candidates are
presented in Paper I, where we also present GC luminosity functions and
global GC specific frequencies of M87 and NGC 4552. The present paper is
dedicated to the investigation of the colour distribution and spatial
distribution of GCs. Observational results are presented in the next
section and are discussed further in \S~\ref{discussion}. We summarize
this paper in \S~\ref{summary}. We adopt a distance of 16.1 Mpc
(distance modulus of 31.03) to M87, and 15.4 Mpc (distance modulus of
30.93) to NGC 4552, based on measurements using the surface brightness
fluctuation method (Tonry et al. 2001). An angular scale of 1$^{\prime}$
corresponds to 4.7 kpc and 4.5 kpc at the distance of M87 and NGC 4552,
respectively.

\begin{figure*}
\begin{center}
 \includegraphics[height=17.5cm,keepaspectratio]{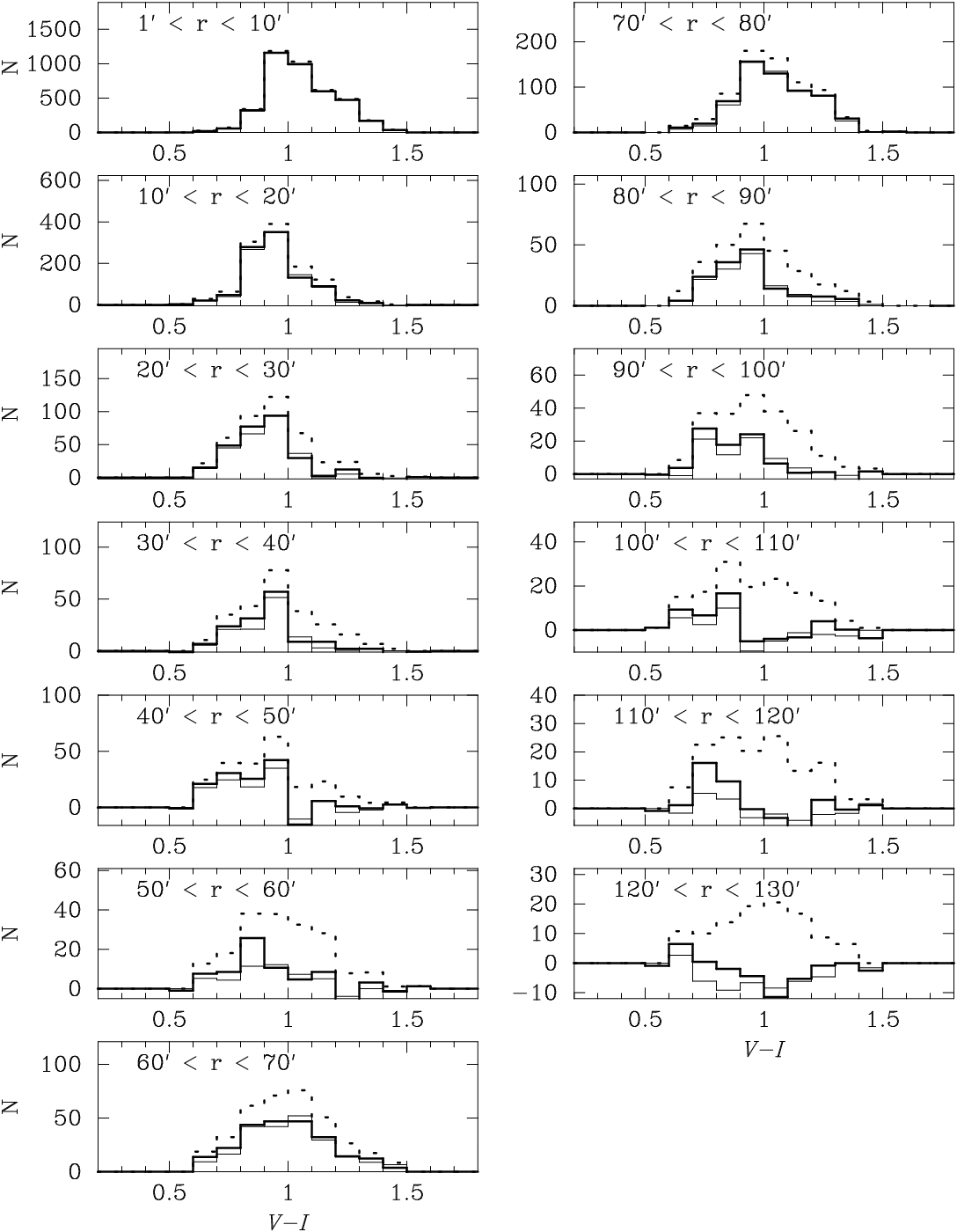}
 \caption{$V-I$ colour distributions of GC candidates with $V \leq 24$
 mag. The dotted line shows the raw colour distribution, and the solid
 lines indicate those after correcting incompleteness and contamination.
 The thick (thin) line shows the colour distribution where the
 subtraction of contaminating sources is performed using the HDF (LH)
 data, respectively.}  \label{allvih}
\end{center}
\end{figure*}

\begin{figure*}
\begin{center}
 \includegraphics[height=16cm,keepaspectratio]{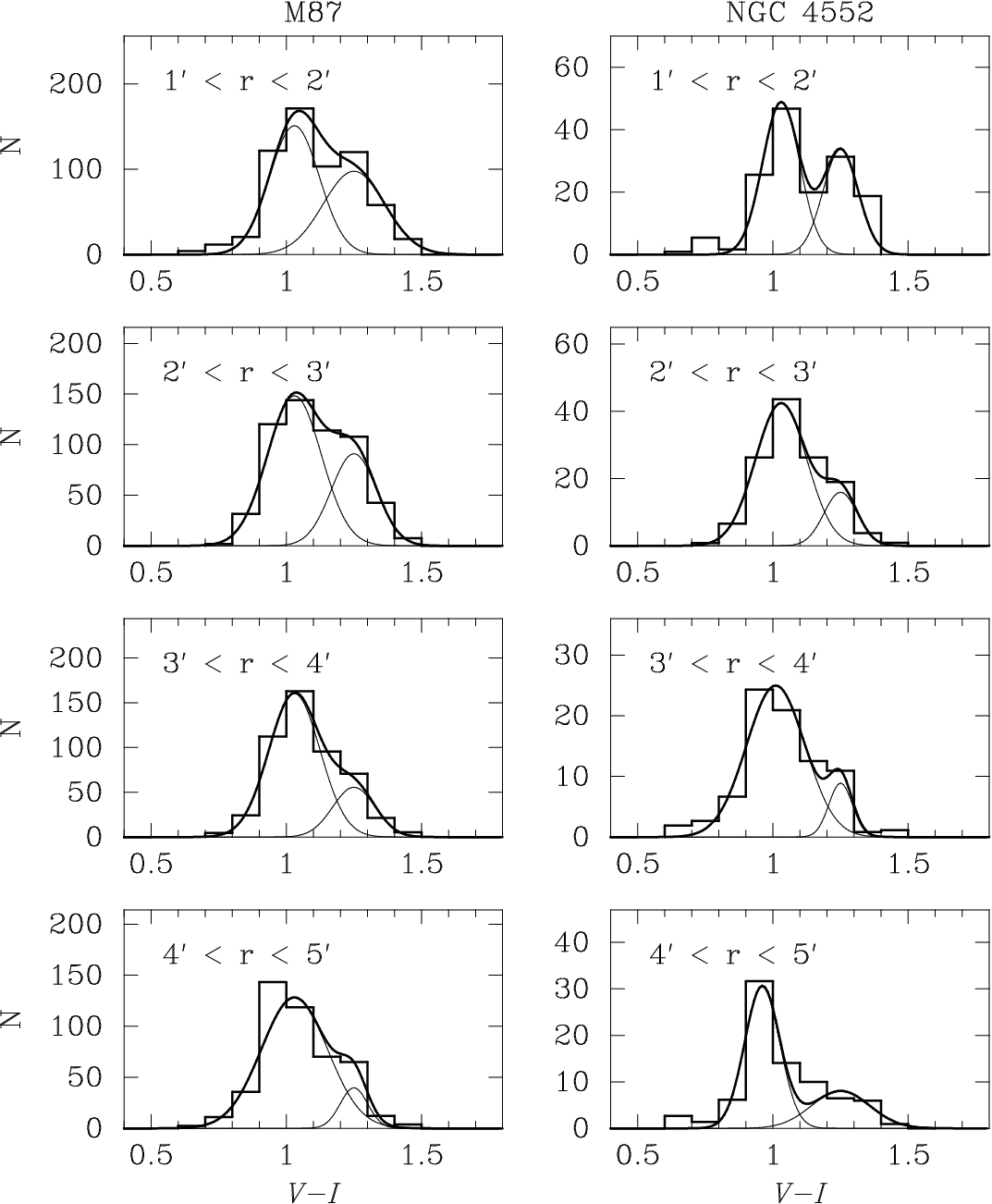}
 \caption{$V-I$ colour distributions for the GC candidates with $V \leq
 24$ mag in the central ($R \leq 5^{\prime}$) regions of M87 ({\it
 left}) and NGC 4552 ({\it right}). The histogram is the GC colour
 distribution obtained after the incompleteness correction and control
 field subtraction are applied. Overplotted are fitted Gaussians
 representing the blue and red GC subpopulations (thin lines) and the
 sum of these (thick lines). A linear relationship between $V-I$ colour
 and metallicity can be derived by using simple stellar population
 models by Vazdekis et al. (1996) with an age of 12.6 Gyr for and a
 Salpeter initial mass function as follows: $[{\rm Fe/H}] = 4.33~(V-I) -
 5.33$.}  \label{bimo}
\end{center}
\end{figure*}

\begin{figure*}
\begin{center}
 \includegraphics[height=13cm,angle=-90,keepaspectratio]{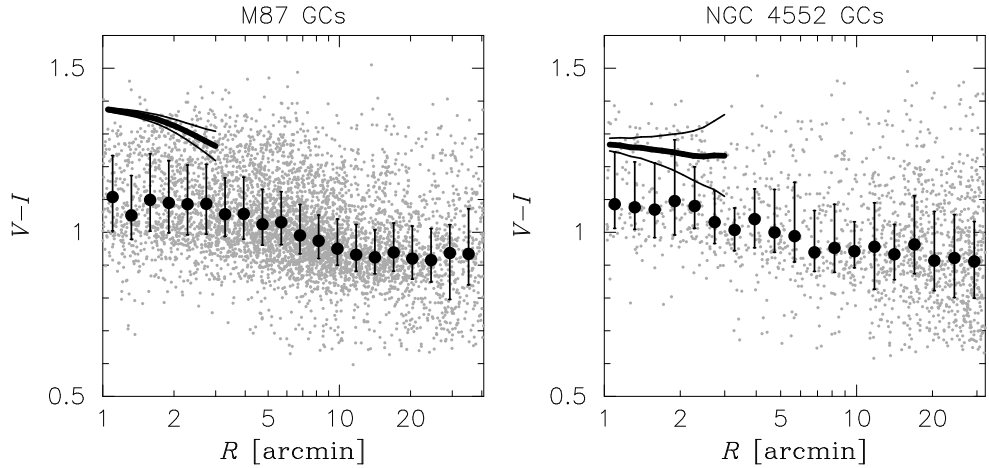}
 \caption{$V-I$ colours of the GC candidates with $V \leq 24$ mag are
 plotted (grey points) against distance from M87 ({\it left panels}) and
 NGC 4552 ({\it right panels}). The average color of GCs at a given
 radius is overplotted with filled circles and the 25- and 75-percentile
 of the colour distribution are indicated with the error bars. The
 colour profiles of the host galaxies are overplotted with solid lines;
 the thick line indicates the measured colour profile and the thin lines
 show the uncertainty due to $\pm 1 \sigma$ errors in the background
 estimation.}
 \label{gcscol}
\end{center}
\end{figure*}

\section{RESULTS}
\label{results}

\subsection{Colour Distribution}
\label{colordist}

Since the errors in the $V-I$ colours tend to be smaller than those of
$B-V$ (see Paper I), we investigate trends of GC colour distributions
with distance from the host galaxy using $V-I$ colours. Also, in order
to keep as much sensitivity as possible to any features in the GC colour
distribution, we restrict the GC sample to those brighter than $V=24$
mag, where incompleteness corrections are not significant (Paper I). In
Fig. \ref{allvih}, GC colour distributions within annuli centered on M87
are shown. The dotted lines indicate raw GC colour distributions without
incompleteness correction and control field subtraction. The thick
(thin) lines show results after incompleteness correction and control
field subtraction using the data from the HDF-N (LH) field,
respectively. Most of these colour distributions show deviations from a
single Gaussian and the sum of two Gaussians give a better fit as
indicated by KMM tests (Ashman, Bird \& Zepf 1994).
However, the presence of a clearly defined red peak at $V-I \sim 1.2$ is
confined primarily to the inner regions of both galaxies. In Fig.
\ref{bimo}, the GC colour distributions within a restricted distance of
$5^{\prime}$ ($\sim 23$ kpc) from each galaxy centre are shown. Double
Gaussians are fitted to these colour distributions and are shown as
smooth curves. In both M87 and NGC 4552, evidence for bimodality is seen
with peaks at $V-I \sim$1.2 mag and $\sim$1.0 mag, respectively, as
previously found in the core regions ($\lesssim 1^{\prime}$) with HST
(Whitmore et al. 1995; Elson \& Santiago 1996a; Elson \& Santiago 1996b;
Neilsen \& Tsvetanov 1999; Kundu et al. 1999; Kundu \& Whitmore 2001;
Larsen et al. 2001; Jord\'{a}n et al. 2002). Using simple stellar
population models by Vazdekis et al. (1996) with an age of 12.6 Gyr and
a Salpeter initial mass function, a linear relationship between $V-I$
colour and metallicity\footnote{Yoon, Yi \& Lee (2006) have recently
suggested a non-linearity in the relationship between GC colour and
metallicity driven by metallicity dependence of horizontal branch
morphology. This would need to be considered when GC metallicity is
discussed in detail, whilst we use the linear approximation in this
paper because almost all the discussions are on a colour basis and do
not depend on calibration of colour to metallicity.} is found to be as
follows:

\begin{equation}
 [{\rm Fe/H}] = 4.33~(V-I) - 5.33. \label{eq:metalvi}
\end{equation}

\noindent
The metallicities corresponding to the peak colours are therefore
estimated to be $[{\rm Fe/H}] \sim -0.13$ and $-1.0$. Fig. \ref{bimo}
also suggests that the bimodality is less clear in the outer regions due
mainly to the decreasing contribution of red GCs (see also Dirsch et
al. 2003; Forbes et al. 2004).\footnote{This trend is similar to the
global trend seen in individual galaxies for which the red GC
subpopulation is less significant in less luminous early-type galaxies
(e.g., Peng et al. 2006). Both may reflect a decrease of average GC
metallicity with decreasing galaxy mass and increasing galactocentric
distance, given the colour-metallicity calibration by Yoon et al.
(2006).} We note that the peak colours of the GC colour distributions in
Fig. \ref{bimo} tend to be slightly ($\sim 0.05$ mag) redder than the
estimates using the HST data corrected for Galactic extinction 
($\Delta (V-I) \sim 0.03$ mag, which is considered in our study). This
difference may be attributed to a calibration error, giving an
uncertainty of $\sim$ 0.2 dex in metallicity estimated from $V-I$. This
discrepancy however does not affect any discussions hereafter.

In Fig. \ref{gcscol}, $V-I$ colours of GC candidates with $V \leq 24$
mag are plotted against distance from M87 and NGC 4552. The average GC
colour is calculated in a radial bin and is indicated with a circle.
The 25- and 75-percentiles of the colour distributions in each radial
bin are shown by the top and bottom of error bars, respectively. This
figure suggests that the mean GC colour becomes bluer with increasing
distances from the host galaxy centres. The colour gradient is estimated
to be $d (V-I) / dr = -0.014 \pm 0.005$ or $d (V-I) / d \log r = -0.17
\pm 0.07$ for the M87 GC population and $d (V-I) / dr = -0.012 \pm
0.006$ or $d (V-I) / d \log r = -0.16 \pm 0.08$ for the NGC 4552 GC
population. These estimations are performed using the GC colours within
15$^{\prime}$ from the host galaxy centre. These colour gradients are
probably due to a combination of the decreasing fraction of red GCs with
radius and a colour gradient within each subpopulation. The latter is
clearly suggested by Fig. \ref{gcscol} at least for the blue GC
subpopulation; whereas $\sim$ 45 \% of the blue GCs ($V-I \leq 1.1$) are
bluer than $V-I = 1.0$ mag near the galaxy centre (e.g., $1^{\prime}
\leq R \leq 2^{\prime}$), the fraction increases (i.e. the average
colour of the blue GCs becomes bluer) with radius and it becomes $\sim$
80 \% at $9^{\prime} \leq R \leq 10^{\prime}$. This is the case both in
M87 and NGC 4552.

However, GC colour gradients seem to be much less significant outside of
$\sim 15^\prime$ ($\sim 70$ kpc) in both of the luminous ellipticals.
The absence of a colour gradient outside a certain radius is
qualitatively consistent with the result for NGC 4472 by Rhode \& Zepf
(2001) who find a marginal colour gradient ($d (B-R) / dr = -0.010 \pm
0.007$) within $8^{\prime}$ or $\sim 40$ kpc, but no significant colour
gradient outside of this radius. These results suggest that the blue GC
population has no metallicity gradient in the outer region of a luminous
elliptical galaxy.
We note that the colour selection of GC candidates is not designed to
sample GCs bluer than $V-I \sim 0.7$, but this colour corresponds to
$[{\rm Fe/H}] \sim -2.3$ if GCs are old (see equation
(\ref{eq:metalvi})) and GCs with this extremely low metallicity are
expected to be rare.

The solid line overplotted in Fig. \ref{gcscol} shows the colour profile
of the host galaxy halo light; the thick line indicates the measured
colour profile and the thin lines show the envelope due to $\pm 1
\sigma$ errors in the background estimation, which is represented by the
standard deviation of pixel values on a CCD frame in the same field as
the galaxy but with no very bright stars or galaxies present.
The colour gradient of the host galaxy halo light is estimated to be $d
(V-I) / d \log r = -0.19 \pm 0.03$ in M87 and $d (V-I) / d \log r =
-0.09 \pm 0.09$ in NGC 4552 within $3^{\prime}$. This indicates that in
M87, the colour gradient of the host galaxy tends to be steeper than the
GC colour gradient, while both are mutually consistent within the errors
for NGC 4552.
We note that the colour gradient of the M87 halo light is steeper than
previous estimates in the core region ($r \lesssim 100^{\prime\prime}$)
e.g. $d (B-R) / d \log r = -0.07 \pm 0.02$ in Peletier et al. (1990) and
$d (B-I) / d \log r = -0.05 \pm 0.01$ in Goudfrooij et al. (1994), so
even less steep gradients are expected in $V-I$ colour.
\footnote{Using simple stellar population models by Vazdekis et al.
(1996) with an age of 12.6 Gyr old and a Salpeter initial mass function,
the sensitivities of $V-I$, $B-R$ and $B-I$ to metallicity variation are
compared as $\Delta(B-R)/\Delta(V-I) = 1.36$ and
$\Delta(B-I)/\Delta(V-I) = 1.92$. These can be used to convert a $B-R$
or $B-I$ colour gradient to that in $V-I$, assuming that colour
gradients are driven by metallicity variations (e.g., Tamura et al.
2000).}
The difference probably comes from the significant bluing of colour with
increasing radius exhibiting in our data especially at $r \gtrsim
120^{\prime\prime}$ (Fig. \ref{gcscol}), which has not been investigated
in the previous studies.
Fig. \ref{gcscol} also indicates that the host galaxy colour, varying at
radii between $1^{\prime}$ and $3^{\prime}$ from 1.37 mag to 1.26 mag in
M87, and 1.27 mag to 1.24 mag in NGC 4552, is slightly redder than the
mean colour of the red GC subpopulation ($V-I \sim 1.2$ mag). We
restrict the host galaxy colour profiles to those inside $3^{\prime}$
because they are highly uncertain outside of this radius.

\subsection{Spatial Distribution}
\label{gcdensity}

\subsubsection{Radial profiles of GC surface densities}

\begin{figure*}
 \begin{center}
  \includegraphics[height=16cm,angle=-90,keepaspectratio]{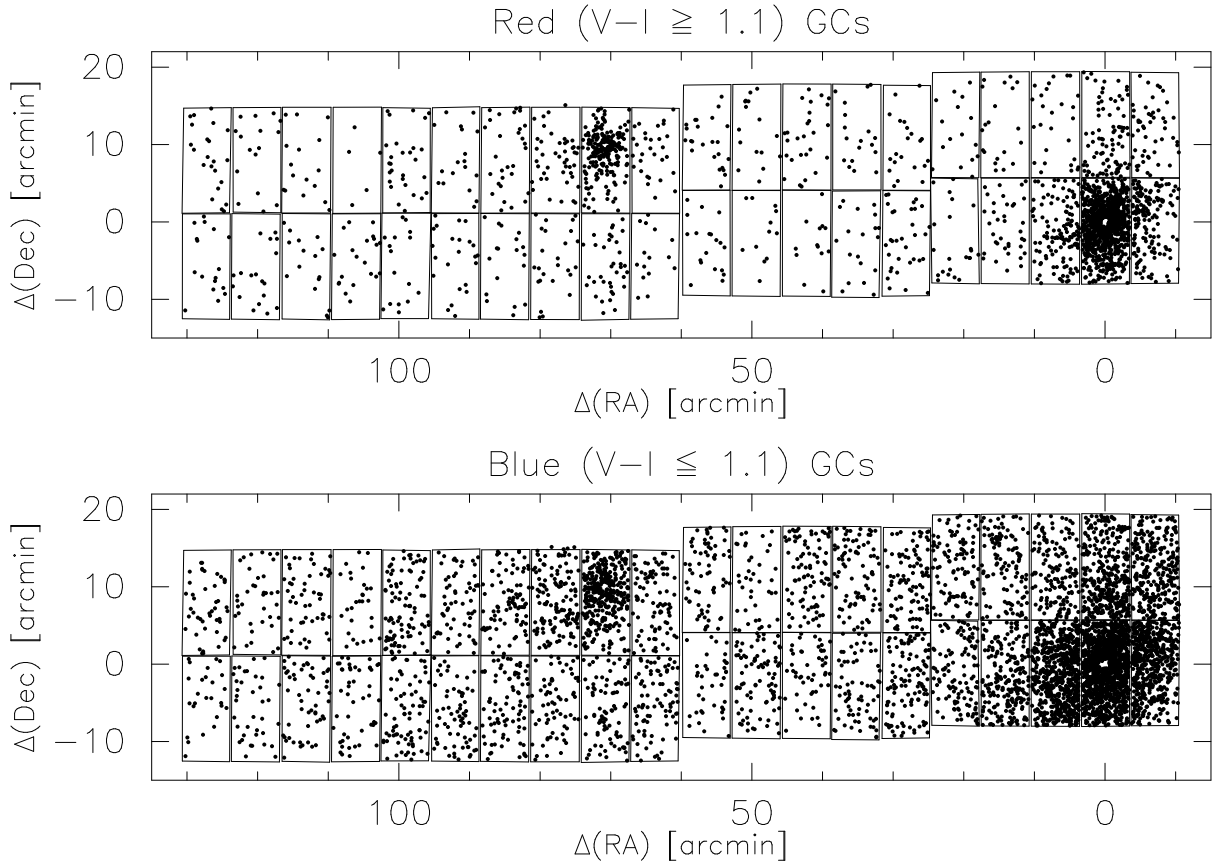}
 \end{center}
 \caption{Spatial distribution of red ($V-I > 1.1$) GCs and blue ($V-I
 \leq 1.1$) GCs. M87 is at coordinates (0, 0), and NGC 4552 is at (72,
 10). Only GCs with $V \leq 24.5$ mag are plotted.}  \label{allmap}
\end{figure*}

The spatial distribution of the GC candidates with $V \leq 24.5$ mag is
shown in Fig. \ref{allmap}. The red GCs ($V-I > 1.1$) are plotted in the
upper panel, and the blue GCs ($V-I \leq 1.1$) are plotted in the lower
panel (the boundary of the colour between red GCs and blue GCs
approximately separates the two peaks in the bimodal colour
distribution). M87 is at coordinates (0, 0), and NGC 4552 is at (72,
10). This figure indicates that the red GCs exist only near the luminous
ellipticals, while the blue GCs tend to be extended out to larger
distances. This trend can be more clearly seen in a plot of GC surface
densities as a function of distance from the host galaxy, which is shown
below.
We use the HDF-N data to subtract the foreground and background
contamination but our results and subsequent discussions do not change
if the LH data are used.

In order to compute the total number of GCs as a function of distance
from the host galaxy (M87 or NGC 4552), we integrate the GCLF in each
radial bin down to either the magnitude where the completeness in that
radial bin is 50 \% or $V = 24.5$ mag whichever is the brighter.
We then multiply the number of GCs by a correction factor to take into
account the GCs fainter than this magnitude limit. This correction
factor is calculated using the Gaussian fit to the GCLF in the inner
regions of M87 and NGC 4552 (Paper I) and it is used independently of
the radial bin. Since the shapes of the GCLFs for the red and blue GC
subpopulations appear to be different (Paper I), the correction factor
for each subpopulation is calculated from the Gaussian fit to the GCLF
of that subpopulation.

\begin{figure*}
\begin{center}
 \includegraphics[height=15cm,angle=-90,keepaspectratio]{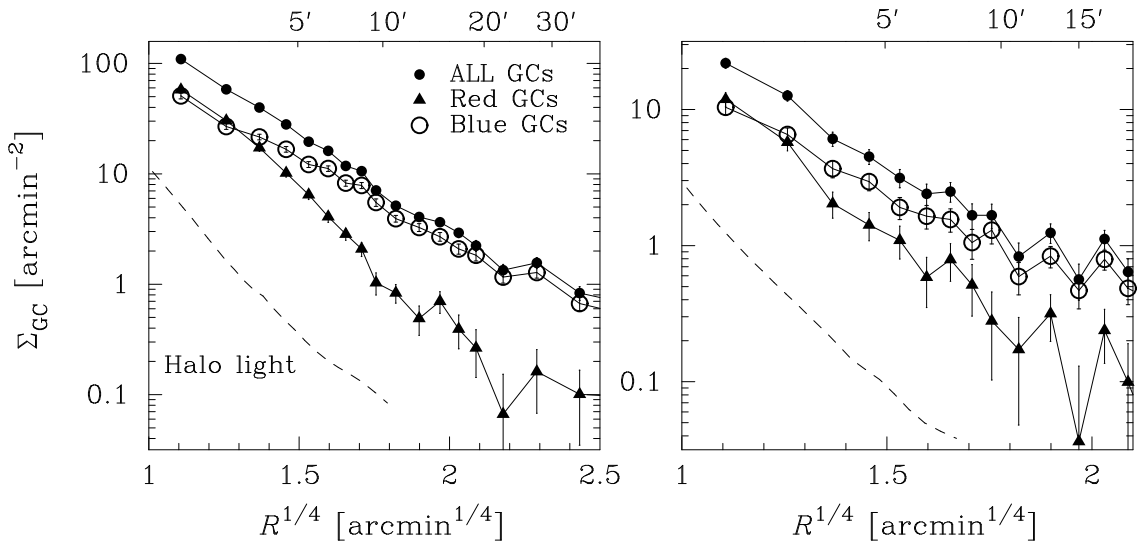}
 \caption{Radial profile of GC surface number densities around M87 ({\it
 left panel}) and NGC 4552 ({\it right panel}). Filled triangles and
 open circles show the surface number densities of the red ($V-I > 1.1$)
 and blue ($V-I \leq 1.1$) GCs, respectively, and solid circles are
 those of the total GC population. Dashed lines show the host galaxy
 halo light distribution in the $V$ band, which is arbitrarily shifted
 in the vertical direction for comparison with the GC profiles.}
 \label{gcsurf1}
\end{center}
\end{figure*}

Radial profiles of GC surface number densities are presented in Fig.
\ref{gcsurf1} for the M87 GCs in the left panel and the NGC 4552 GCs in
the right panel. Filled triangles and open circles are the number
densities of the red and blue GC subpopulation, respectively, and solid
circles are those of the total GC population. The error bar represents a
Poisson error at a given radius estimated from the number of GCs in this
annulus and the number of contaminating objects in the control field
normalized to the same area. 
The slopes of these radial profiles are calculated by fitting regression
lines. For M87, the slope of the red, blue, and total GC density profile
($d \log \Sigma_{\rm GC} / d R^{1/4}$) is $-2.44 \pm 0.06$, $-1.47 \pm
0.03$, and $-1.76 \pm 0.03$, respectively. The slopes for the NGC 4552
GC profiles are quite similar to those for the M87 GCs; $-2.44 \pm
0.16$, $-1.60 \pm 0.10$, and $-1.88 \pm 0.08$ for the red, blue, and
total GC density profile, respectively.
The host galaxy halo light distributions in the $V$ band are also shown
by dashed lines. These are arbitrarily shifted in the vertical direction
to enable comparison with the GC distributions.
The slope of the surface brightness distribution ($d \log I_{V} / d
R^{1/4}$) is calculated to be $-2.36 \pm 0.09$ for M87 and $-2.38 \pm
0.11$ for NGC 4552.

Figure \ref{gcsurf1} therefore demonstrates that the GC distribution
tends to be more extended than the host galaxy halo light, which has
also been found in NGC 4472 (Harris 1986), NGC 1399 (Dirsch et al.
2003), NGC 4649 (Forbes et al. 2004) and NGC 4374 (G\'{o}mez \& Richtler
2004). It also indicates that while the distribution of the red GC
subpopulation is as centrally concentrated as (or possibly slightly more
concentrated than) that of the halo light, the distribution of the blue
GC subpopulation tends to be more extended as has been pointed out
previously not only for M87 (C\^{o}t\'{e} et al. 2001) but also for some
other luminous ellipticals: NGC 4472 (Lee, Kim \& Geisler 1998), NGC
4649 (Forbes et al. 2004) and NGC 1399 (Bassino et al. 2006). We
investigate the GC distribution in further detail using the full radial
coverage of our data in \S~\ref{models}.
In contrast to these results,
no clear difference is found in another luminous Virgo elliptical, NGC
4374 (G\'{o}mez \& Richtler 2004). This may indicate that the spatial
distribution of GCs varies
from galaxy to galaxy. It should be mentioned, however, that the control
field used in the G\'{o}mez \& Richtler (2004) study is relatively close
to the host galaxy;
$10^{\prime}$ ($\sim$ 53 kpc) from NGC 4374, where a number of blue GCs
may still exist. Contamination of the control fields by blue GCs would
tend to steepen the radial profile of the blue GC surface densities,
making it more similar to that of red GC subpopulation.

\subsubsection{Local GC Specific frequency}

\begin{figure*}
\begin{center}
 \includegraphics[height=15cm,angle=-90,keepaspectratio]{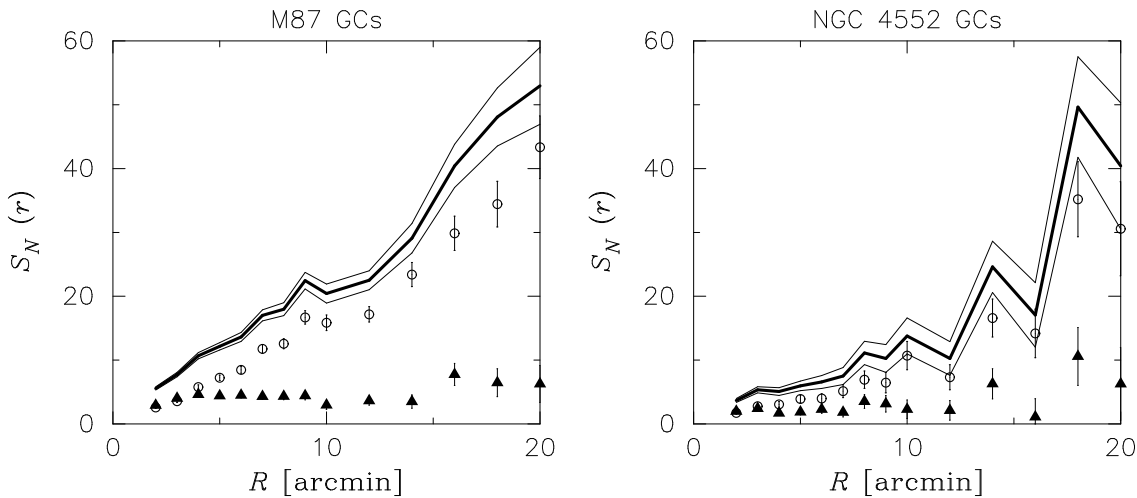}
 \caption{Local GC specific frequency ($S_N$) is plotted against
 distance from the host galaxy. The thick line shows the measured values
 and the thin lines show the envelope of the $\pm 1\sigma$ uncertainty
 in the estimation of the total number of GCs. $S_N$ for the red GC
 subpopulation is shown as filled triangles, and that for the blue GC
 subpopulation is indicated with open circles.}  \label{snfig}
\end{center}
\end{figure*}

In order to convert the radial profiles of GC surface densities to those
of local GC specific frequency ($S_N$),
the luminosity of host galaxy halo light within an annulus is calculated
using a de Vaucouleurs law fitted to the $V$-band surface brightness
distribution measured using our data (Fig. \ref{gcsurf1}). The results
are shown in Fig. \ref{snfig}. The solid line indicates the measured
$S_N$ variation with radius and the thin lines show the 1 $\sigma$
uncertainty according to the errors in the GC surface densities. We also
calculate the $S_N$ values for only the red GCs or blue GCs shown by
filled triangles and open circles, respectively.
This figure indicates that the increasing trend of local $S_N$ with
distance from the host is primarily due to the fact that the blue GC
distribution tends to be more extended than the host galaxy halo light
distribution.
As suggested by the similarity of the red GC distribution to the halo
light distribution, there is no clear trend of $S_N$ with radius for the
red GCs. This is the case both in M87 and NGC 4552.

\section{DISCUSSION}\label{discussion}

\subsection{Detailed Analysis of GC Surface Density Profiles} \label{models}

\subsubsection{Host of the blue GCs}

The extended nature of the blue GCs compared to the host galaxy halo
light distribution may imply the existence of a number of GCs which are
not associated with any luminous galaxies.
In order to examine the validity of an idea that i-GCs exist in a galaxy
cluster (e.g., White 1987; Muzzio 1987), which could apparently enhance
the number of GCs in a luminous elliptical galaxy and its $S_N$ by
superposition, West et al. (1995) predict the local number densities of
i-GCs at the locations of 14 giant cluster ellipticals by assuming that
i-GCs are distributed in a galaxy cluster following the mass
distribution of the cluster and that a more massive galaxy cluster
possesses a larger number of i-GCs (therefore the apparent enhancement
of GC number due to i-GCs would depend both on the mass of the host
cluster and the location of the galaxy in the cluster). They find that
higher i-GC densities are predicted for those giant ellipticals with
larger excess numbers of GCs compared to the predictions from a typical
$S_N$ value for normal ellipticals and claim that this correlation is
evidence for the existence of such i-GCs.

\begin{figure*}
\begin{center}
 \includegraphics[height=15cm,angle=-90,keepaspectratio]{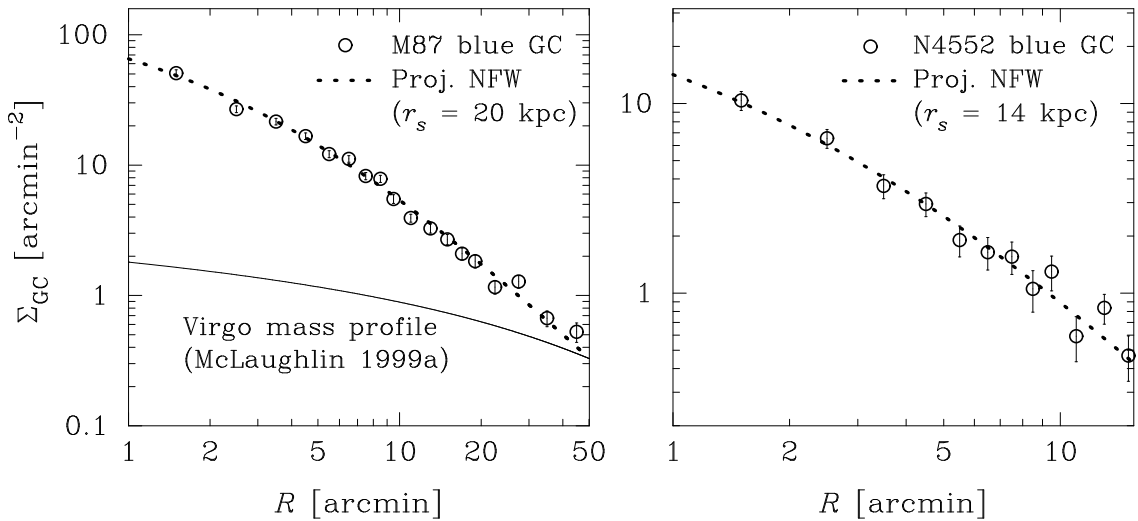}
 \caption{{\it Left panel:} Surface densities of the M87 blue GCs are
 plotted with open circles against logarithmic distance from the centre
 of M87. The dotted line indicates a projected NFW profile with a scale
 radius of 20 kpc 
 which fits the blue GC distribution. The solid line indicates the
 radial dependence of the projected Virgo cluster mass distribution
 represented by 
 an NFW profile with a scale radius of 560 kpc (McLaughlin 1999a). {\it
 Right panel:} Same as the left panel, but for the NGC 4552 blue
 GCs. The horizontal axis is logarithmic distance from the centre of NGC
 4552. The dotted line shows a projected NFW profile with a scale radius
 of 14 kpc.}
 \label{bluexray}
\end{center}
\end{figure*}

Harris et al. (1998) argue against this idea by comparing the radial
surface density profile of all the M87 GCs within $10^{\prime}$ from the
centre with that of the GCs around NGC 4472, which has a similar
luminosity to M87 but is located in the outer region of the Virgo
cluster and has a normal $S_N$ value. They demonstrate that the excess
number of GCs in M87 compared to that in NGC 4472 is seen at all radii,
in the sense that an additional GC population to explain the excess
would need to be highly concentrated around M87, which is unlikely for
an i-GC population. This argument still holds based on more recent GC
data (this study for M87 GCs and Rhode \& Zepf (2001) for NGC 4472 GCs).
Our data also suggest that while the blue GC distribution around M87 is
more extended than the halo light distribution (Fig. \ref{gcsurf1}), it
is significantly more concentrated toward the galaxy centre compared to
the radial profile of the dark matter surface mass density in the Virgo
cluster derived by McLaughlin (1999a), which is indicated in the left
panel of Fig. \ref{bluexray}. The cluster mass distribution in this case
is described by an NFW profile $\rho (r) \propto (r/r_s)^{-1}(1 +
(r/r_s))^{-2}$ (Navarro, Frenk \& White 1997), where $r_s = 560$ kpc,
that is constrained over a large range of distance from the cluster
centre by the stellar mass distribution of M87 (de Vaucouleurs \& Nieto
1978), the surrounding X-ray hot gas distribution (Nulsen \&
B\"{o}ringher 1995), the surface density profile of dwarf elliptical
galaxies (Binggeli, Tammann \& Sandage 1987), and the kinematics of
Virgo early-type galaxies (Girardi et al. 1996). The solid line in the
left panel of Fig. \ref{bluexray} shows the 2D projection of this
NFW profile\footnote{In this radial range, this model exhibits a radial
profile steeper than other formulae used to represent the cluster mass
distribution such as a King profile fitted to the galaxy distribution
(nearly constant with radius; Binggeli et al. 1987) and that determined
by the GC kinematics within $\sim 8^{\prime}$ ($\sim 40$ kpc) of M87
($\propto R^{-0.3}$; Cohen \& Ryzhov 1997; C\^{o}t\'{e} et
al. 1998). The surface brightness distribution in the faint envelope of
M87 ($\gtrsim 15^{\prime}$) is however similar to this projected
Hernquist profile (Carter \& Dixon 1978; see also Harris et al. 1998).}
with the normalization scaled arbitrarily for comparison with the GC
profile and can be seen to be significantly more extended than the blue
GC distribution.
It should also be pointed out that such an extended distribution of blue
GCs is seen not only around M87 but also NGC 4552 (Fig. \ref{gcsurf1}),
NGC 4472 (Lee et al. 1998) and NGC 4649 (Forbes et al. 2004) which are
located in the outer regions of the Virgo cluster. Both of these facts
suggest that the contribution of i-GCs existing on the cluster scale is
insignificant around M87 and most of the blue GCs as well as the red GCs
must be associated with the host galaxy.

\subsubsection{Blue GC distribution and host galaxy dark matter halo}

The next question would then be what the blue GC distribution
represents. One interesting possibility is that the blue GCs are
associated with the dark matter halo of the host galaxy itself.
There are at least some suggestions that dark matter halos of luminous
(massive) ellipticals are more spatially extended than their stellar
contents from gravitational lens analyses of early-type galaxies at
intermediate redshifts (Rusin, Kochanek \& Keeton 2003; Treu \& Koopmans
2004; Ferreras, Saha \& Williams 2005) and X-ray hot gas analyses of
nearby ellipticals (Fukazawa et al. 2006). Our data suggest that the
blue GC distributions around M87 and NGC 4552 can be modelled by NFW
profiles. 
In Fig. \ref{bluexray}, the dotted lines show projected NFW profiles
with scale radii ($r_s$) of 20 kpc and 14 kpc fitted to the blue GC
distribution of M87 and NGC 4552, respectively. The 1 $\sigma$ fitting
errors in $r_s$ are estimated to be 1 kpc for M87 and 3 kpc for NGC 4552
by Monte Carlo simulations.

Now we examine whether the extents of the blue GC distributions are
consistent with those of the dark matter halos of M87 and NGC 4552. It
is difficult to investigate the dark matter distribution of M87 itself
because it is confused with the mass distribution of the Virgo cluster;
the continuously rising velocity dispersion of the M87 GCs out to large
radii
implies that the Virgo cluster potential already dominates at $\sim$
5$^{\prime}$ $\sim$ 23 kpc (Romanowsky \& Kochanek 2001).
Meanwhile, there is no study so far of the dark matter profile of NGC
4552. Hence, we will refer to investigations of other galaxies and
compare the blue GC distributions with estimates of the dark matter
halos of galaxies with comparable luminosities to M87 and NGC 4552.

From weak lensing analysis of field galaxies, Hoekstra, Yee \& Gladders
(2004) derived an average scale radius in the NFW profile of
23$^{+5}_{-4}$ kpc for a galaxy with $L_B = 2 \times 10^{10} L_{B,
\odot}$, which corresponds to $M_B = -20.3$ mag. If we use the same
scaling relation $r_s \propto L_{B}^{0.75}$ as adopted by Hoekstra et
al. (2004), which is suggested by the observed relations among galaxy
mass, luminosity and velocity dispersion, the scale radii of the dark
matter halos in M87 and NGC 4552 can be calculated. Since the $B$-band
absolute magnitude of M87 (NGC 4552) is $-21.6$ mag ($-20.2$ mag),
respectively, the scale radii would be 56$^{+12}_{-10}$ kpc for M87 and
21$^{+5}_{-4}$ kpc for NGC 4552. The blue GC distributions may therefore
be less extended than the dark matter halos of these ellipticals,
although the discrepancy is much less significant for NGC 4552.
Alternatively, cluster galaxies may have more compact dark matter halos
than field galaxies (Natarajan et al. 1998; Limousin, Kneib \& Natarajan
2006). This environmental dependence could be expected due to, e.g.,
earlier collapse epochs of dark matter halos in denser environments
resulting in more compact mass density profiles (e.g., Diemand, Madau \&
Moore 2005) and/or effects of tidal interactions in clusters which can
strip off the outer part of a galaxy halo (e.g., Bullock et al. 2001;
Avira-Reese et al. 2005). Observational evidence for the tidal stripping
scenario has been presented by Natarajan, Kneib \& Smail (2002), who
estimated sizes of dark matter halos of $L^{\ast}$ galaxies in several
rich clusters at intermediate redshifts by analysing galaxy lensing
signals on the HST/WFPC2 images and found that the size of a galaxy halo
with a certain velocity dispersion depends on the mass density of the
galaxy cluster.
While both of these effects may have been at work for NGC 4552, the
latter would be less likely for M87 because tidal stripping is
ineffective for a massive galaxy located at the centre of a galaxy
cluster. Further investigations of GC populations and mass profiles of
ellipticals both in the field and clusters will be needed to see whether
the blue GC distribution is as extended as the dark matter halo.

\begin{figure*}
\begin{center}
 \includegraphics[height=15cm,angle=-90,keepaspectratio]{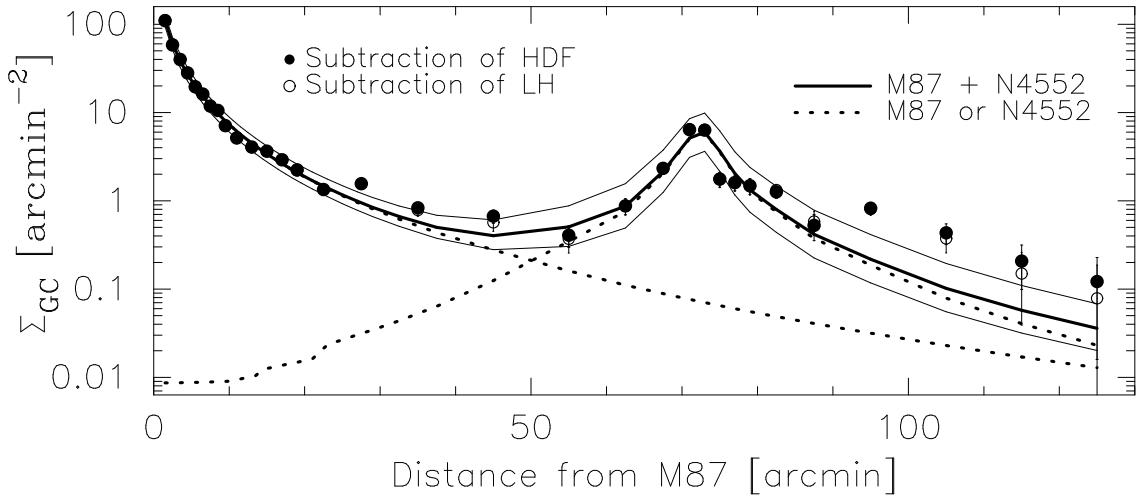}
 \caption{GC surface number density is plotted over the full range of
 distance from M87 covered by our data. The GC densities calculated
 using the HDF-N data for subtraction of contamination are indicated
 with solid circles and those using the LH data are plotted with open
 circles. The dotted lines indicate the radial profiles of GC surface
 densities associated with M87 and NGC 4552, and the sum of the two
 distributions is indicated with the thick solid line. The thin solid
 lines around this thick line show the 1 $\sigma$ envelope due to the
 fitting errors of the de Vaucouleurs laws to the measured GC densities
 around M87 and NGC 4552.} \label{full}
\end{center}
\end{figure*}

\subsubsection{Is there any evidence for intergalactic GCs?}

Given that the majority of the blue GCs as well as the red GCs appear to
belong to the host galaxy, is there any clear evidence for an
intergalactic GC (i-GC) population in the GC spatial distribution?
In Fig. \ref{full}, the GC surface densities for the whole coverage of
our data are plotted as a function of distance from M87. The dotted line
near M87 represents the GC distribution associated with M87 obtained by
fitting a de Vaucouleurs law to the radial profile of total GC surface
densities within $20^{\prime}$ and extrapolating to larger distances.
This is repeated for NGC 4552 using the GC profile within $10^{\prime}$
of the centre and the dotted line peaking around 75$^{\prime}$ from M87
shows this component. The thick solid line shows the sum of these two GC
distributions. The thin solid lines around this thick line show the 1
$\sigma$ envelope due to the fitting errors of the de Vaucouleurs laws
to the measured GC densities around M87 and NGC 4552. This plot
indicates that the GC surface densities around $100^{\prime}$ from M87
most clearly exceed those expected from just the M87 GCs and NGC 4552
GCs, suggesting that an additional i-GC population may exist with a
surface density of $\sim$ 0.2 arcmin$^{-2}$. Our data do not exclude the
possibility that i-GCs exist not only around $100^{\prime}$ from M87 but
at all radii, provided the surface density is less than $\sim 0.1$
arcmin$^{-2}$.

The estimated i-GC surface density depends sensitively on the assumed
radial distributions of GCs associated with M87 and NGC 4552. For
example, the GC distributions may not follow the de Vaucouleurs laws and
any deviations from this assumption would be a source of uncertainty in
the existence and amount of possible i-GCs. Even if the de Vaucouleurs
laws are good approximations of the {\it galactic} GC distributions, we
need to consider the fitting errors. When we adopt the upper end of the
envelope (i.e. $+1 \sigma$) as the galactic GC distribution, the excess
GC number density would be reduced down to $\sim$ 0.1 arcmin$^{-2}$. On
the contrary, the galactic GCs may not be distributed out to such large
radii and their distribution may be truncated at a certain distance from
the host galaxy. If this is the case, all the GCs found at distances
larger than $90^{\prime}$ from M87 could be i-GCs and the number density
would be increased up to $\sim 0.5$ arcmin$^{-2}$.
While these systematic effects dominate the uncertainty of the i-GC
surface number density, the effect of foreground and background
subtraction error is another concern. The 1 $\sigma$ Poisson error
calculated from the numbers of GC candidates and contaminating objects
is $\sim$ 0.1 arcmin$^{-2}$. In addition, there is a difference of
$\sim$ 0.05 arcmin$^{-2}$ in the surface density of contaminating
objects between HDF-N and LH, which is presumed to be due to cosmic
variance of the two control fields and/or difference of data
quality. Therefore the GC surface density of 0.2 arcmin$^{-2}$ estimated
at a distance of $115^{\prime}$ from M87 is a detection at a $1.5 -
2~\sigma$ level of significance, although this significance level
depends on the adopted model for the galactic GC distributions.

In the following discussions, we will adopt 0.2 arcmin$^{-2}$ as the best
estimate of the i-GC number density from our survey with the lower (upper)
limit of 0.1 (0.5) arcmin$^{-2}$, respectively. These numbers are
tentative and need to be confirmed with future observations.
Spectroscopy would be a promising way to identify true i-GCs in the
candidates and their line-of-sight velocity distribution would be
evidence that they are not associated with any luminous galaxies. Since
the i-GCs are expected to be associated with the Virgo cluster
potential, their velocity distribution should be similar to the dynamics
of the Virgo member galaxies which have a dispersion $\sigma$ $\sim$
$500 - 800$ km s$^{-1}$ at $\sim$ 110$^{\prime}$ from M87, depending on
the morphological types of the galaxies used in the calculation
(Binggeli, Sandage \& Tammann 1985). In fact, Arnaboldi et al. (2004)
measured radial velocities for a dozen Virgo intracluster PNe (i-PNe)
and found $\sigma = 1000 \pm 210$ km s$^{-1}$.
More accurate determinations of i-GC surface densities would allow one
to address whether the distribution of the i-GC population in the Virgo
cluster is just a function of distance from M87, or is inhomogeneous.
This could be compared with the recent study of the i-PNe population in
the Virgo cluster, which shows a significant field-to-field variation
and no clear radial trend in the surface number density (Feldmeier et
al. 2004a).

\subsection{Comparison with Studies of Intergalactic PNe: Constraints on
$S_N$ for Intergalactic GC populations}\label{comp}

Although the i-GC surface number density estimated in the previous
section is preliminary,
it provides a useful upper limit. The surface number density of i-PNe
has recently been estimated in several fields of the Virgo cluster
(Feldmeier et al. 2004b; Aguerri et al. 2005) and the associated amount
of intracluster stellar population in the Virgo cluster has been
inferred by means of stellar evolution theory. This enables one to
calculate an expected surface number density of i-GCs using $S_N$ as a
free parameter, provided that i-GCs and i-PNe follow a similar spatial
distribution (this needs to be confirmed in future surveys of both i-GCs
and i-PNe in the same area of sky). In the following, we derive an $S_N$
for the i-GC population by comparing the observed and predicted i-GC
number densities.
While several sources of large uncertainty are expected to exist in this
calculation, it gives a useful insight into which galaxy populations
furnish intracluster stars and GCs, because $S_N$ is believed to be
correlated with galaxy type (e.g., Harris 1991).

In order to infer an i-GC surface number density from the distribution
of i-PNe, we start from the bolometric surface luminosity density of
intracluster stellar light which is estimated using the luminosity
function of PNe to be $5.5 \times 10^5 - 1.0 \times 10^6~L_{\odot}$
kpc$^{-2}$ where detected (Feldmeier et al. 2004b). The lower bound
value corresponds to $1.2 \times 10^7~L_{\odot}$ arcmin$^{-2}$.
Assuming a bolometric correction for intracluster stars of $-0.8$
(Jacoby, Ciardullo \& Ford 1990; Feldmeier et al. 2004b), this can
further be transformed to a surface density of $V$-band luminosity of $V
\sim -12.7$ mag arcmin$^{-2}$. The predicted i-GC surface number density
is therefore $\sim 0.07 \times S_N$ arcmin$^{-2}$. According to our
analyses of the GC surface number density profiles, the i-GC surface
number density is $\lesssim$ 0.2 arcmin$^{-2}$ where they are likely to
exist. Therefore, the results from i-GCs and i-PNe would be mutually
consistent if $S_N \sim 2.9$ for the i-GC population.
Considering that GCs tend to be more extended than halo stars and may be
more easily tidally stripped than stars, the $S_N$ value of 2.9 seem to
be unexpectedly low. However, our understanding of such extended GC
distributions comes mainly from observations of luminous galaxies and GC
distributions in less luminous early-type galaxies and spiral galaxies,
which are the more common galaxy populations in clusters, are much less
well studied.
The $S_N$ for i-GCs could also depend on the type of galaxy cluster;
since the Virgo cluster is a spiral-rich cluster, a significant amount
of stars and GCs may have been provided by spiral galaxies whose $S_N$
is low ($\sim 1$) resulting in a modest average $S_N$ for i-GCs.

There are a number of sources of uncertainty in $S_N$ for i-GC
populations included in the above derivation. One is obviously the i-GC
surface density. If we adopt 0.1 (0.5) arcmin$^{-2}$ as the i-GC surface
density (see the last section), the $S_N$ would be $\sim$ 1.4 (7.1),
respectively. In addition, the luminosity function of i-PNe and the
bolometric corrections for intracluster stellar populations are not very
well constrained and their uncertainties are expected to be added to
those already accounted for. The $S_N$ of i-GC populations will
therefore need to be updated by forthcoming GC and PNe surveys.

\subsection{Implications for the Origin of GC Populations}
\label{formation}

In \S~\ref{models}, we suggest that most of the blue GCs around M87 are
associated not with the Virgo cluster but with the host galaxy, based on
the fact that the spatial distribution of blue GCs around M87 is not as
extended as the Virgo cluster mass distribution, and that such an
extended distribution of blue GCs is not a special characteristic of
M87. 
The extended distribution of blue GCs could therefore be an intrinsic
property of the host galaxy and this dependence of spatial distribution
on GC colour (metallicity) should therefore be explained by models of GC
and elliptical galaxy formation. In addition, recent observations
suggest that the spectroscopic ages of both metal-rich and metal-poor
GCs in ellipticals are as old as those of Milky Way GCs (Strader et al.
2005a) and that the mean colours of the red GCs and the blue GCs are
both correlated with host galaxy properties (Larsen et al. 2001;
Strader, Brodie \& Forbes 2004; Strader et al. 2005b; Peng et al.
2006). We now examine the existing scenarios of GC and elliptical galaxy
and try to constrain the physical processes involved with all of these
recent observatios considered.

In the multiphase collapse scenario, the blue GC subpopulation is
considered to have formed {\it in situ} during a dissipative collapse of
the protogalactic cloud at high redshift but this occurred
before the bulk of the halo stars and red GCs formed (Forbes et al.
1997). All GCs and their host elliptical galaxy would form within a
short time scale and GCs would therefore be uniformly old. Since GC
formation is presumed to be affected by the depth of the potential well
of the host galaxy, the average metallicities of metal-rich and
metal-poor GCs should correlate with the host galaxy properties. This
scenario also appears to be supported by the possible evidence for a
metallicity gradient in the blue GCs (\S~\ref{colordist}), suggesting
that the GC formation was accompanied by a dissipative collapse of the
protogalactic gas cloud. The absence of a metallicity gradient at the
large distances from the host galaxy centre does not necessarily
contradict such a dissipative process; it may be due to the presence of
lower limit in GC metallicity at $[{\rm Fe/H}] \sim -2$ (Cayrel 1986;
Burgarella, Kissler-Patig \& Buat 2001; Woodley, Harris \& Harris 2005;
Harris et al. 2006).
However, the physical mechanism for this {\it in situ} formation to
produce metal-poor and metal-rich GCs remains unclear. Numerical
simulations seem to indicate that when a massive elliptical galaxy forms
at high redshift following the dissipative collapse of a protogalactic gas
cloud, most of the stars (and probably GCs as well) formed in a single
episode of intense star formation (e.g., Chiosi \& Carraro 2002), which
is unlikely to create separate metal-rich and metal-poor GC
subpopulations.
Even if the starburst could be split into two phases, still missing
would be clues to explain why GCs rather than halo stars are formed
predominantly in the early phase of star formation and why the
distribution of these metal-poor GCs is more extended than that of the
metal-rich GCs.

In fact, these problems could be resolved at least to some extent by
incorporating some effects of galaxy mergers. Instead of mergers of
mature spirals as originally proposed by Ashman \& Zepf (1992), here we
consider galaxy mergers accompanied by starbursts and red GC formation
to be complete at high redshift, since no clear evidence for red GCs to
be significantly younger than blue GCs has so far been found from
observations (Cohen, Blakeslee \& Ryzhov 1998; Strader et al. 2005a).
These high redshift mergers could play a key role in the formation of a
massive elliptical galaxy and its metal-poor GC subpopulation as
follows: (1) Around the peak of the primordial density fluctuations
where a massive elliptical galaxy will subsequently form, sub-galactic
clumps collapse first; (2) these sub-galactic clumps are presumably
gas-rich and stars and GCs form therein, both of which are expected to
be metal-poor; (3) when these sub-galactic clumps merge into a massive
elliptical galaxy, the metal-poor GCs also assemble and result in a
substantial fraction of the blue GC subpopulation of the host galaxy
(see also discussions by Harris et al. 2006).
The metal-poor GCs are expected to behave as collisionless particles in
this assembly and their spatial distribution when assembled could
therefore follow a mass distribution of the host galaxy at this time.
Since this assembly is expected to take place in the dark matter
potential of a massive elliptical galaxy, the distribution of metal-poor
GCs should be similar to a dark matter distribution of the host galaxy
and explain why an NFW profile can give a good fit to the blue GC
subpopulation.  Note that the presence of a metallicity gradient in the
blue GCs might constrain such a dissipationless assembly of the
metal-poor GCs. The average metallicity of the blue GCs should correlate
with the host galaxy mass if a more massive elliptical galaxy forms
through an assembly of more massive building blocks, which are expected
to possess more metal-rich GCs as suggested for nearby dwarf ellipticals
(Lotz, Miller \& Ferguson 2004). Meanwhile, the bulk of field stars and
red GCs are presumed to have formed in the starbursts following
dissipative mergers of these gas-rich sub-galactic clumps. Their spatial
distributions could therefore be similar to each other and more compact
than that of the blue GCs due to the effect of dissipation.

On the other hand, the high $S_N$ values observed in giant ellipticals
located at the cluster centres would need to be explained by biased GC
formation in denser (presumably more gas-rich) environments (West 1993;
Blakeslee 1999; McLaughlin 1999b). Rhode, Zepf \& Santos (2005)
investigate GC populations in several ellipticals and spirals and find
that the number of blue (metal-poor) GCs normalized by the stellar mass
of the host galaxy tends to increase with the host galaxy stellar mass,
which may also be evidence that formation of metal-poor GCs was more
biased around more massive galaxies.
In the high redshift merger scenario proposed above, this correlation
could be explained if a more massive elliptical formed from sub-galactic
clumps with a higher GC formation efficiency. In this regard, however,
mergers of more massive sub-clumps to form a more massive elliptical
appear to be implausible because less luminous local dwarf ellipticals
tend to have higher $S_N$ (Durrell et al. 1996; Miller et al. 1998).
Also, it is unclear if the dominant metal-poor GCs assembled in this way
would be found in spiral galaxies.
Another process to form metal-poor GCs surrounding more massive galaxies
independent of galaxy morphology therefore seems necessary. One
candidate may be additional GC formation triggered by the cosmological
reionization
(Cen 2001; see also discussions by Rhode et al. (2005) and references
therein). More theoretical work and numerical experiments will be needed
to predict the GC properties and their evolution within a framework of
structure and galaxy formation for these proposed mechanisms of GC
formation to be compared with the observations quantitatively.

Finally, we comment on the presence of tidally stripped (blue) GCs from
other (less luminous) galaxies found preferentially in the outer regions
of luminous ellipticals.
In galaxy clusters, the tidal force from the cluster potential is
expected to act on galaxies to remove halo stars and GCs most
efficiently around the core radius of a galaxy cluster (e.g., Merritt
1984), which is 45 kpc ($\sim 10^{\prime}$) for the Virgo cluster
(Nulsen \& B\"{o}hringer 1995). The GC surface density around a central
cluster galaxy like M87 may therefore be enhanced around this radius by
tidal capture. For luminous ellipticals located in the outer region of a
galaxy cluster or in even less dense environments, only galaxy-galaxy
interactions could perhaps be at work. Detailed predictions of the
spatial distribution of GCs expected from tidal interactions would be
useful to place constraints on the accretion scenario. Our estimate of
$S_N$ for i-GCs is another observational constraint on the validity of
this tidal capture model around M87 and other giant ellipticals with
high $S_N$ values. If the estimated $S_N$ for the i-GC population
represents the true ratio of GCs to tidally stripped stars, the $S_N$ of
a central galaxy would be expected to decrease as the number of
interactions increases, which is perhaps hard to be reconciled with the
observed high $S_N$ for the M87 GCs.

\section{SUMMARY AND CONCLUSION}\label{summary}

We have performed a wide-field imaging survey of the globular cluster
(GC) populations around M87 with Suprime-Cam on the 8.2m Subaru
Telescope. A $2^{\circ} \times 0_{\cdot}^{\circ}5$ (560 kpc $\times$ 140
kpc) field extending from M87 to the east was observed through the $BVI$
filters. GC candidates are selected not only with an extended source cut
but also with a colour selection which includes almost all the Galactic
GCs but minimizes the contamination of foreground stars and background
galaxies. We also analyze archival imaging data of the control fields
(HDF-N and Lockman hole) to assess foreground and background
contamination in the GC candidates which needs to be statistically
subtracted from the GC candidates.

In this paper, we investigate the colour distribution, surface number
density, and local specific frequency ($S_N$) of GC candidates as a
function of distance from two luminous elliptical galaxies M87 and NGC
4552. In investigating these statistical properties, we subtract
foreground and background contamination using both the HDF-N data and
the Lockman Hole data but we find that the results do not depend
significantly on choice of the control fields. The main results are
summarized as follows:

 \begin{itemize}

  \item The colour distributions of GCs in the innermost region (within
	$5^{\prime}$ or $\sim 20$ kpc from the galaxy centre) shows
	bimodality both in M87 and NGC 4552. This bimodality becomes
	less clear in the outer regions due to a decreasing contribution
	of the red GC subpopulation. The colour of the host galaxy halo
	light is slightly redder ($\Delta(V-I) \sim 0.1$ mag) than that
	of the red GC subpopulation.

  \item The average colour of GCs becomes bluer with increasing distance
	from the host galaxy. The colour gradient within 15$^{\prime}$
	from the host galaxy is estimated to be $d (V-I) / d \log r =
	-0.17 \pm 0.07$ for the M87 GC population and $d (V-I) / d \log
	r = -0.16 \pm 0.08$ for the NGC 4552 GC population. The GC
	colour gradient may be due to a combination of the decreasing
	contribution of red GCs and a colour gradient in each GC
	subpopulation. At radii larger than $\sim 15^{\prime}$, however,
	the radial gradient of GC colour seems much less significant in
	both of M87 and NGC 4552. The colour gradient of the host galaxy
	halo light is found to be steeper than the GC colour gradient in
	M87, while both are consistent with each other in NGC 4552.

  \item The spatial distribution of the GC population as a whole is more
	extended than the host galaxy halo light both in M87 and NGC
	4552. This is primarily due to the extended distribution of the
	blue GCs, while the distribution of the red GC subpopulation is
	as centrally concentrated as that of the halo light.
	Accordingly, the local $S_N$ for the blue GCs increases with
	radius, whereas that for the red GCs is nearly constant.

  \item Although the spatial distribution of blue GCs around M87 is more
        extended than that of the host galaxy halo light and the red
	GCs, it appears to be less extended than the mass distribution
	of dark matter in the Virgo cluster. Furthermore, the extended
	distribution of blue GCs does not seem to be a special
	characteristic of giant ellipticals at the cluster centres like
	M87. Both of these facts do not support the idea that the high
	$S_N$ value of M87 is due to the superposition of intergalactic
	GCs (i-GCs). Instead, we suggest that most of the blue GCs as
	well as the red GCs are associated with the host galaxy.

  \item We find marginal evidence for the existence of intracluster GCs
	(i-GCs) with a surface density of $0.2^{+0.3}_{-0.1}$
	arcmin$^{-2}$. Comparison with the predicted surface luminosity
	density of intracluster stars gives a crude estimate of $S_N$
	for this i-GC population $S_N = 2.9^{+4.2}_{-1.5}$. If this
	$S_N$ represents the true ratio of GCs to tidally stripped
	stars, the fraction of GCs captured by this process would be low
	in the GC population of M87.

\end{itemize}

\section*{ACKNOWLEDGEMENTS}

We are grateful to the referee, Dr. Terry Bridges, for careful reading
of our manuscript and for insightful comments which significantly
improved this paper.
This work was based on data collected at Subaru Telescope operated by
National Astronomical Observatory of Japan and those obtained from the
SMOKA system operated by the Astronomical Data Center, National
Astronomical Observatory of Japan .
We appreciate the members of the Subaru Telescope operation team,
especially Dr. Hisanori Furusawa for supports during the observation.
This work was partly supported by Grants-in-Aid for Scientific Research
(Nos. 16540223 and 17540216) by the Japanese Ministry of Education,
Culture, Sports, Science and Technology.

\label{lastpage}

\end{document}